\documentclass[useAMS,usegraphicx,usedcolumn,usenatbib]{mn2e}

\def\prd{Phys.~Rev.~D} 
\def\apj{ApJ} 
\def\apjs{ApJS} 
\def\mnras{MNRAS} 
\def\aap{A\&A} 

\def\VEV#1{{\left\langle #1 \right\rangle}}
\def\alm{a_{\ell m}}
\def\dlm{d_{\ell m}}
\def\nlm{n_{\ell m}}
\def\wlm{w_{\ell m}}
\def\Mll{M_{\ell \ell'}}
\def\Dl{D_\ell}
\def\Cl{C_\ell}
\def\l#1{\ell_#1}
\def\m#1{m_#1}
\newcommand{\wjjj}[6]
{{
\left( 
\begin{array}{lcr} #1 & #2 & #3 \\#4 & #5 & #6 \end{array}
\right) 
}}
\newcommand\order[1] {${{\cal O}\! \left( #1 \right)}$}

\title[Xspect]{ Xspect, estimation of the angular power spectrum by computing cross-power spectra with analytical error bars}

\author[M.~Tristram, J.~F.~Mac{\'\i}as-P\'erez, C.~Renault, D.~Santos]{
  M.~Tristram, J.~F.~Mac{\'\i}as-P\'erez, C.~Renault, D.~Santos\\
  Laboratoire de Physique Subatomique et de Cosmologie, 53 Avenue des Martyrs, 38026 Grenoble Cedex, France}

\begin{document}

\maketitle

\date{\today}

\begin{abstract}
  We present Xspect, a method to obtain estimates of the angular power
  spectrum of the Cosmic Microwave Background (CMB) temperature
  anisotropies including analytical error bars developed for the {\sc
    Archeops} experiment. Cross-power spectra are computed from a set
  of maps and each of them is in itself an unbiased estimate of the
  power spectrum as long as the detector noises are uncorrelated.
  Then, the cross-power spectra are combined into a final temperature
  power spectrum with error bars analytically derived from the
  cross-correlation matrix.
  
  This method presents three main useful properties : (1) no
  estimation of the noise power spectrum is needed, (2) complex
  weighting schemes including sky covering and map noise properties
  can be easily taken into account, and corrected for, for each input
  map, (3) error bars are quickly computed analytically from the data
  themselves with no Monte-Carlo simulations involved.  Xspect also
  permits the study of common fluctuations between maps from different
  sky surveys such as CMB, Sunyaev-Zel'dovich effect or mass
  fluctuations from weak lensing observations.
\end{abstract}



\section[]{Introduction}
The measurement of the angular power spectrum of the CMB anisotropies,
$\Cl$s, has become one of the most important tools in modern
cosmology. As long as they remain in the linear regime, the
fluctuations predicted by most inflationary scenarii
\citep{hu,linde,liddle} lead to Gaussian anisotropies on the CMB. Thus
the angular power spectra in temperature and polarization contain all
the cosmological information on the CMB sky. Cosmological parameters
and other physical quantities of interest in the early Universe can be
directly derived from them. In parallel to the explosion of CMB
datasets both in size and quality (WMAP \citep{wmap}, {\sc Archeops}
\citep{archeops_cl}, {\sc Boomerang} \citep{boom}, {\sc Maxima}
\citep{maxima}, DASI \citep{dasi}, VSA \citep{vsa}, CBI \citep{cbi},
ACBAR \citep{acbar}), fast codes have been developed to estimate the
CMB angular power spectrum (CMBFAST
\citep{cmbfast1,cmbfast2,cmbfast3}, CAMB \citep{camb}) allowing us to
compare fast and efficiently theory and observations using powerful
statistical tests (CMBEASY \citep{cmbeasy}, COSMOMC \citep{cosmomc},
\citep{douspis}). Furthermore, huge efforts are undertaken to ease
the estimation of the angular power spectrum from input CMB maps in
order to cope with larger, deeper and more complex sky surveys in a
reasonable amount of computing time.

Excluding very specific methods -- for example those which are under
study for the {\sc Planck} satellite mission and which take advantage
of the {\sc Planck} ring scanning strategy
\citep{vanLeeuwen,challinor,ansari} -- most CMB power spectrum
estimators can be grouped into two categories~: maximum likelihood and
`pseudo'-$\Cl$ estimators. A complete review and comparison between
the two methods can be found in \citealp{efstathiou}, here we just
discuss the key points of each of them.

Maximum likelihood methods for temperature anisotropies
\citep{bond,tegmark,borrilla} are based on the maximization of the
quadratic likelihood. The method estimates the sky angular power
spectrum from the angular correlation function of the data. Error bars
for the power spectrum are generally computed directly from the
likelihood function which is either fully sampled in the range of
interest or approximated by a quadratic form. Dealing with an
inhomogeneous coverage of the sky involves a great computational
complexity \mbox{\order{N^3_{pix}},} where $N_{pix}$ is the number of
pixels of the input map. Therefore, these methods are very CPU time
consuming for current large datasets like WMAP and probably not well
adapted \citep{borrillb} for future satellite missions like {\sc
  Planck} which will produce maps of the sky of more than $N_{pix} = 5
\times 10^7$ pixels. A generalization of these methods to the analysis
of CMB polarization is discussed in \citealp{tegmark_pol}.

Alternatively, the so-called `pseudo-$\Cl$'s estimators compute
directly the {\it `pseudo'} angular power spectrum from the data.
Then, they correct it for the sky coverage, beam smoothing, data
filtering, pixel weighting and noise biases. A comprehensive
description of this method was first given by \citealp{peebles} and an
application to the angular clustering of galaxies can be found in
\citealp{peebleshauser}. More recently, several approaches to this
method have been developed. Among them, SPICE \citep{spice} and its
extension to polarization \citep{chon} compute in the real space
first the correlation function to correct for the sky coverage bias
and then the power spectrum from the latter. A pseudo-$\Cl$ estimator
in the spherical harmonic space applied to CMB experiments is given in
\citealp{wandelt01pseudo} and \citealp{master} (MASTER). They computes
directly the power spectrum before correct it for the different
biases. An approach applied to apodised regions of the sky is
presented in \citealp{hansen02} and extended to polarization in
\citealp{hansen03}. These estimators can be evaluated using fast
spherical harmonic transforms \order{N^{3/2}_{pix}} and therefore
provide fast and accurate estimates of the $\Cl$s. However, they
require an accurate knowledge of the instrumental setup and noise in
order to correct them for the biases discussed previously. In fact,
they use an estimation of the power spectrum of the noise in the map,
generally computed via Monte-Carlo simulations, which is subtracted
from the original power spectrum. This is also used to estimate the
error bars in the power spectrum by calculating the variance of the
$\Cl$s over the set of simulations.

In this paper, we describe a method to estimate the $\Cl$s by
computing the cross-power spectra between a collection of input maps
coming either from multiple detectors of the same experiment or from
different instruments. The `pseudo' cross-power spectra are explicitly
corrected for incomplete sky coverage, beam smoothing, filtering and
pixelization. Assuming no correlation between the noise contribution
from two different maps, each of the corrected cross-power spectra is
an unbiased estimate of the $\Cl$s. Analytical error bars are derived
for each of them. The cross-power spectra, that do not include the
classical {\it auto}-power spectra, are then combined using a Gaussian
approximation of the likelihood function. In the same way, we can also
compute the estimate of the common angular power spectrum,
$\Cl^{common}$, of sky maps from different experiments.
\\

A similar method also based on the combination of a set of cross-power
spectra has been first used to obtain recent results from the first
year WMAP data \citep{wmap_cl}. The main difference between the method
presented is this paper and the WMAP one is the determination of the
cross-correlation matrix (see~Sect.\ref{correlation}) of the corrected
cross-power spectra used both for the combination of these into a
single power spectra and for the estimation of the error bars on the
latter. The WMAP team estimates the cross-correlation matrix from a
model of the data. This includes specific terms related to the WMAP
data such as the contribution from point sources and the uncertainties
on the beam window functions as well as a term related to the CMB
anisotropies which is estimated from a fiducial model. The WMAP
cross-correlation matrix, used for the combination of the cross-power
spectra, does not incorporate the effects of mode coupling. In a
further step, they account for the mode coupling and the dependence on
the fiducial CMB model for the computation of the uncertainties on the
final power spectrum.

By contrast, the method presented here computes the cross-correlation
matrix directly from the cross-power spectra estimated from the data.
This allows us to include naturally the mode coupling in this matrix.
Further, this permits the computation of analytical error bars (as
described in Sect.~\ref{correlation}) which are very compatible with
those obtained from simulations (see Sect.~\ref{archeops}). Because
of the above, this method can be applied without modification to the
estimation of the power spectrum of the correlated signal between a
set of maps of the sky coming from multiple instruments with
potentially different sky coverages. For example, we have used this
method on {\sc Archeops} data for the estimation of the CMB angular
power spectrum and the contribution from foregrounds to this one
\citep{archeops_cl2}. We have also used it for the estimation of the
foreground emission at the sub-millimeter and millimeter wavelength by
cross-correlating the {\sc Archeops} data with foreground dust
templates.

In Sect.~\ref{cross}, we remind to the reader the computation of the
cross-power spectra from `pseudo' cross-power spectra. In
Sect.~\ref{correlation}, we specify the correlation between
cross-power spectra and between multipoles. Analytical expressions for
the error bars and the covariance matrix for each cross-power spectra
are derived. Section~\ref{combination} discusses the combination of
the cross-spectra from either a single full data set
(Sect.~\ref{lincomb}) or several independent experiments
(Sect.~\ref{common}). Finally, Xspect is applied to simulations of the
{\sc Archeops} balloon-borne experiment in Sect.~\ref{archeops}.

\section[]{Cross-power spectra}\label{cross}
Under the assumption of uncorrelated noise between detectors, the
cross-power spectrum computed from sky maps is a non noise-biased
estimate of the angular power spectrum. In general, when computing the
$\Cl$s, other instrumental effects as beam smoothing, incomplete sky
coverage and time ordered data filtering need to be taken into
account. These effects and the way to correct them have been deeply
covered in the literature for classical `pseudo-$\Cl$' estimators
\citep{spice,master,hansen02} which are noise biased as they use
directly the auto power spectrum of each detector map. As shown in the
following, these corrections can be extended to the case of
cross-power spectra.

The CMB temperature anisotropies, $\Delta T$, over the full-sky can be
decomposed into spherical harmonics as follows,
\begin{equation}
\Delta T(\vec{n}) = \sum_{\ell m} \alm Y_{\ell m}(\vec{n})
\end{equation}
where the coefficient $\alm$ are given by
\begin{equation}
\alm = \int \Delta T (\hat{n}) Y_{\ell m}(\hat{n}) d\Omega \,\,\, .
\end{equation}

The CMB temperature field $\Delta T$ predicted by most inflationary
models \citep{liddle,linde,hu} is in general Gaussian distributed so
that the ensemble average of the $\alm$ coefficients are
\begin{eqnarray}
\VEV{ \alm } & = & 0 \\
\VEV{ \alm a_{\ell' m'}^* } & = & \VEV{\Cl} \delta_{\ell
  \ell^\prime} \delta_{m m^\prime} \label{def_cl}
\end{eqnarray}

An unbiased estimate of the CMB temperature power spectrum $\VEV{\Cl}$
is therefore given by
\begin{equation}
  \widehat{\Cl} = \sum_{m=-\ell}^{\ell}
  \frac{\left| \alm \right|^2}{2 \ell+1} \, \, .
\end{equation}

Ground-based and balloon-borne CMB experiments present an
inhomogeneous sky coverage. On the other hand, satellite experiments
like COBE, WMAP and {\sc Planck}, although they provide full-sky maps,
residuals of foreground contamination in the Galactic plane and point
sources contamination make impossible to use the complete maps when
computing the CMB angular power spectrum. Furthermore, for most CMB
experiments, the noise properties vary considerably due to different
redundancies between pixels of the same map. So obtaining an estimate
of the power spectrum requires a differential weighting of pixels
within the same map which translates into an effective inhomogeneous
sky coverage.

The decomposition in spherical harmonics of the observed temperature
anisotropies including weighting can be written for a single detector
as follows
\begin{eqnarray}
  \dlm
  & = &
  \int d\Omega \,
  T^{\rm map}(\hat n) W(\hat n) Y_{\ell m}(\hat n) \\
  & \simeq &
  \sum_{p=1}^{N_{\rm pix}}
  \Omega_p \, T^{\rm map}_p\, W(\hat n_p)\, Y_{\ell m}(\hat n_p),
\end{eqnarray}
where $\Omega_p = \frac{4\pi}{N_{\rm pix}}$ for equal pixels (as in
HEALPix pixelization, \citealp{healpix}) and $W$ is the mask applied
to the input sky temperature $T^{map}$. This temperature can be
decomposed into signal and noise, $T^{map} = T^{signal} + T^{noise}$
which are assumed to be uncorrelated, {\it i.e.}
\begin{equation}
  \VEV{T^{signal}\,T^{noise}} = 0 \,\,.
\end{equation}

The effect of a non-homogeneous coverage of the sky can be described
in spherical harmonics \citep{peebles} by a mode-mode coupling
matrix $\Mll$ which depends only on the angular power spectrum of the
weighting scheme (hereafter weighting mask) applied to the sky to
account for the incomplete coverage, the removal of the Galactic plane
and the inhomogeneous noise properties of the detector.

Thus, the `pseudo-$\Cl$' estimator $\Dl$ is defined as follows
\begin{equation}
  \widehat{\Dl}
  =
  \frac{1}{2 \ell+1}
  \sum_{m=-\ell}^{\ell}
  \left| \dlm \right|^2 \, \, .
\end{equation}

The relation that links the `pseudo' power spectrum, directly measured
 on the sky, and the power spectrum $\Cl$ of the CMB
anisotropies is given by
\begin{equation}
  \widehat{\Dl}
  =
  \sum_{\ell'} \Mll |p_{\ell'} B_{\ell'}|^2 F_{\ell'}
  \VEV{C_{\ell'}}+\VEV{N_\ell}
  \label{pseudo_cl}
\end{equation}
where $B_\ell$ is the beam transfer function describing the beam
smoothing effect; $p_\ell$ is the transfer function of the
pixelization scheme of the map describing the effect of smoothing due
to the finite pixel size and geometry; $F_\ell$ is an effective
function that represents any filtering applied to the time ordered
data; and $\VEV{N_\ell}$ is the noise power spectrum.

\subsection[]{Pseudo cross-power spectrum}
An unbiased estimate of the cross-power spectrum $C_\ell^{AB}$ between
the full sky maps of two independent and perfect detectors $A$ and $B$
can be obtained from
\begin{equation}
  \widehat{C_{\ell}^{AB}} = 
  \frac{1}{2 \ell+1}
  \sum_{m=-\ell}^{\ell} a_{\ell m}^A\,a_{\ell m}^{B\,*}.
\end{equation}
where $a_{\ell m}^A$ and $a_{\ell m}^B$ are the coefficients of the
spherical harmonic decomposition of maps $A$ and $B$ respectively.

In the same way, we can compute the `pseudo' cross-power spectrum
$\widehat{\Dl^{AB}}$ between any two detectors $A$ and $B$ by
generalizing Eq.~\ref{pseudo_cl}
\begin{equation}
  \widehat{\Dl^{AB}}
  =
  \sum_{\ell'} \Mll^{AB} |p_{\ell'}|^2 B_{\ell'}^A B_{\ell'}^B F_{\ell'}^{AB}
  \VEV{C_{\ell'}^{AB}}+\VEV{N_\ell^{AB}} \, \, \, .
  \label{pseudo_cross}
\end{equation}
Each of the terms in Eq.~\ref{pseudo_cross} is described in more
details in the following subsections.

\subsection[]{Noise cross-power spectrum, $N_\ell^{AB}$}
The main advantage of using cross-power spectra is that the noise is
generally uncorrelated between different detectors
\begin{equation}
  \VEV{\nlm^A n_{\ell^\prime m^\prime}^{B\,*}} = 0 \,\,\, . \nonumber
\end{equation}
This assumption will be maintained throughout this paper.

Thus, the cross-power spectra are straightforward estimates of the
angular power spectrum on the sky and for two different detectors
({\it i.e.} $A \ne B$), the `pseudo' cross-power spectrum reads
\begin{equation}
  \widehat{\Dl^{AB}}
  =
  \sum_{\ell'} \Mll^{AB} |p_{\ell'}|^2 B_{\ell'}^A B_{\ell'}^B F_{\ell'}^{AB}
  \VEV{C_{\ell'}^{AB}} \, \, \, .
  \label{pseudo}
\end{equation}

\subsection[]{Coupling kernel matrix, $\Mll^{AB}$}\label{coupling_kernel}
The coupling kernel matrix $\Mll$, introduced in Eq.~\ref{pseudo_cl}
and described in details in \citealp{master}, reads
\begin{equation}
  \Mll = \frac{2\ell'+1}{4\pi}\sum_{\ell''}
  (2\ell''+1) {\mathcal{W}}_{\ell''} \wjjj{\ell}{\ell'}{\ell''}{0}{0}{0}^2
  \label{mll_master}
\end{equation}
where $\mathcal{W}_\ell=\frac{1}{2\ell+1}\sum_{m=-\ell}^{\ell}\left|
 \wlm \right|^2$ is the power spectrum of the mask. It takes into
account the mask applied to the data where mask represents both the
sky coverage and the weighting scheme.

Equation~\ref{mll_master} can be easily generalized to the case of two
different masks applied respectively to each map of the two detectors
involved in the cross-power spectrum calculation. Replacing the
quadratic terms $|\wlm|^2$ by $\VEV{\wlm^A\wlm^{B*}}$ in the
computation of Eq.~\ref{mll_master} leads to
\begin{equation}
  \Mll^{AB} = \frac{2\ell'+1}{4\pi}\sum_{\ell''}
  (2\ell''+1) {\mathcal{W}}^{AB}_{\ell''} \wjjj{\ell}{\ell'}{\ell''}{0}{0}{0}^2
\end{equation}
where 
$\mathcal{W}_\ell^{AB}=\frac{1}{2\ell+1}\sum_{m=-\ell}^{\ell}\wlm^A
\wlm^{B*}$, the cross-power spectrum of the masks.

This property allows us to deal with independent masks representing
different sky coverages and to apply an appropriate specific weighting
scheme to each detector map. Note that the correction in the multipole
space discussed here is fully equivalent to an appropriate
normalization of the cross correlation between the two sky masked maps
in real space. This analogy is important as it helps to understand why
no fully overlapping masks for the input maps can be considered.

\subsection[]{Filter function, $F_\ell^{AB}$}
The filter function $F_\ell$ accounts for the filtering of the time
ordered data which is generally needed in most CMB experiments either
to avoid systematic effects or to reduce correlated low frequency
noise. The time domain filtering is performed along a preferred
direction on the sky (scanning direction) and so leads commonly to an
anisotropic sky even if the assumption of initial isotropic
temperature fluctuations holds. In this case, the estimates of the
angular power spectrum provided by Eq.~\ref{pseudo_cl} and
Eq.~\ref{pseudo} are not exact any more and should be corrected for a
function both in $\ell$ and $m$, $F_{\ell,m}$. Obtaining accurate
estimates of such a correction is particularly complex and for most
cases, as proposed by \citealp{master}, the correction for an effective
$F_{\ell}$ is good enough for the accuracy required in the
reconstruction of the CMB power spectrum. The $F_{\ell}$ function can
be, for example, computed via Monte-Carlo simulations of the sky from
which mock time ordered data are produced for each of the detectors
involved and then filtered.

From an initial theoretical CMB power spectrum, we compute a large
number of realizations of the sky using the HEALPix software {\it
 synfast} \citep{healpix} and compute mock time ordered data from the
scanning strategy of each detector. Maps are then computed with and
without filtering before re-projection. The $F_\ell$ function is
obtained from the mean ratio of the `pseudo' power spectra of the
filtered and not filtered maps. The latter are obtained using the
HEALPix software {\it anafast}.

In the case of the cross-power spectra and considering the previous
approximation, an effective filter function $F_\ell^{AB} =
\sqrt{F_\ell^A F_\ell^{B}}$ will be considered in the following. As
defined, the effective filtering function allows us to consider
detectors for which the time domain filtering is different.

Note that, for nearly white noise and all-sky surveys such as WMAP or
{\sc Planck} missions, filtering may not be required and thus $F_\ell=1$.

\subsection[]{Beam window function, $B_\ell^A$}
The beam window function $B_\ell$ describes the smoothing effect of
the main instrumental beam under the hypothesis of circularity. The
latter does not hold in general as for most experiments the main beam
pattern is asymmetric. As beam uncertainties have become the most
important source of systematic errors, taking into account the
asymmetry of the beam pattern is necessary. Several solutions have
been proposed either circularizing the beam \citep{wu,page} or
assuming an elliptical Gaussian beam \citep{souradeep,pablo}. The work
of \citealp{wandelt01beam} presents how to convolve {\it exactly} two
band limited but otherwise arbitrary functions on the sphere - which
can be the 4-$\pi$ beam pattern and the sky descriptions. An analytic
framework for studying the leading order effects of a non-circular
beam on the CMB power spectrum estimation is proposed in
\citealp{mitra}.

The authors of this paper present {\it Asymfast} \citep{asymfast}, a
general method to estimate an effective $B_\ell$ function taking into
account the asymmetry of the main beam and the scanning strategy. {\it
  Asymfast} is based on a decomposition of the main beam pattern into
a linear combination of Gaussians which permits fast convolution in
the spherical harmonic space along the scanning strategy.

\subsection[]{Cross-power spectrum from `pseudo' cross-power spectrum}
The cross-power spectrum $C_{\ell}^{AB}$ can be obtained from the
`pseudo' cross-power spectrum $D_{\ell}^{AB}$ by resolving
Eq.~\ref{pseudo_cross} which leads to invert the coupling kernel
matrix ${\Mll}^{AB}$. In general for complex sky coverage and
weighting schemes, this matrix is singular and can not be inverted
directly. To avoid this problem, \citealp{master} proposed the binning of
the coupling kernel matrix which reduces considerably the complex
correlation pattern in the $\Cl$s introduced by the applied mask. The
binning is obtained by applying the operators $P_{b\ell}$ and $Q_{\ell
 b}$ as follows
\begin{eqnarray}
  \widehat{C_b^{AB}} & = & P_{b\ell} \widehat{C_\ell^{AB}} \\
  \widehat{D_b^{AB}} & = & P_{b\ell} \widehat{\Dl^{AB}} \\
  Q_{\ell b} & = & P_{b\ell}^{-1}.
\end{eqnarray}

The solution of Eq.~\ref{pseudo} in the new base reads

\begin{equation}
  \widehat{C_b^{AB}} =  {\cal M}_{bb'}^{-1} \widehat{D_{b'}^{AB}}
\end{equation}
with
\begin{equation}
  {\cal M}_{bb'} = P_{b\ell}
  \left( \Mll^{AB} F_{\ell'} p_{\ell'}^2 B_{\ell'}^A B_{\ell'}^B \right)
  Q_{\ell' b'}.
\end{equation}

As the correlation between multipoles $\ell$ depends very much on the
instrumental setup, the sky coverage and the weighting scheme, the
binning has to be defined for each experiment. It is a compromise
between a good multipole sampling and low correlations between
adjacent bins.

Hereafter to avoid confusion and makes the notation simpler all
equations will be written in $\ell$ instead of $b$.

\section[]{Cross-correlation matrix - analytical error bars and covariance matrix}\label{correlation}
From $N$ input maps we can obtain $N(N-1)/2$ cross-power spectra
$\Cl^{AB}$ ($A \neq B$) which are unbiased estimates of the angular
power spectrum but which are obviously not independent. In this
section, we describe the estimation of the cross-correlation matrix
between cross-spectra and between multipoles. We show how the error
bars and the covariance matrix in multipole space can be deduced for
each cross-power spectra.

\subsection[]{The cross-correlation matrix, $\Xi$}\label{correlation_matrix}
Given a sky map from a detector $A$, it can be combined to each of the
other detector maps to form $N-1$ cross-power spectra which will 
therefore be highly correlated. Furthermore, due to the masking we also
expect that each cross-power spectra will be correlated for adjacent
multipoles and thus, correlations between adjacent multipoles will be
also present between different cross-power spectra. To describe this
complexity we define the cross-correlation matrix
$$\Xi^{AB,CD}_{\ell \ell'} = \VEV{
  \left( C_\ell^{AB}- \VEV{C_\ell^{AB}}\right)
  \left( C_{\ell'}^{CD}-\VEV{ C_{\ell'}^{CD}}\right)^*}$$
of the cross-power spectra $AB$ ($A \ne B$) and $CD$ ($C \ne D$) which
can be fully computed as shown in the following.

From Eq.~\ref{pseudo} we can express the `pseudo' cross-power spectrum
between detectors $A$ and $B$ as follows
\begin{eqnarray}
  \widehat{\Dl^{AB}}
  & = &
  {\cal M}_{\ell \ell'}^{AB} \widehat{C_{\ell'}^{AB}}
\label{estimator}
\end{eqnarray}
where ${\cal M}_{\ell \ell'}^{AB} = \Mll^{AB} E_{\ell'}^{A}
E_{\ell'}^{B}$ and $E_\ell =p_\ell B_\ell \sqrt{F_\ell}$ and therefore
the corrected cross-power spectrum for detectors $A$ and $B$ is given
by
\begin{equation}
  \widehat{C_{\ell}^{AB}} = \left({\cal M}_{\ell \ell'}^{AB} \right)^{-1} \widehat{D_{\ell'}^{AB}}.
\end{equation}

Using the above expression and following the previous definition, the
cross-correlation matrix $\Xi^{AB,CD}_{\ell \ell'}$ reads
\begin{eqnarray}
  \Xi^{AB,CD}_{\ell \ell'}
  & \equiv &
  \VEV{ \Delta C_\ell^{AB} \Delta C_{\ell'}^{CD^*} }
  \nonumber \\
  & = &
  \VEV{ {\cal M}_{\ell\ell_1}^{AB\,-1} \Delta D_{\ell_1}^{AB}
    \left( {\cal M}_{\ell'\ell_2}^{CD\,-1} \Delta D_{\ell_2}^{CD}\right)^* }
  \nonumber \\
  & = &
  {\cal M}_{\ell \ell_1}^{AB\,-1}
  \VEV{\Delta{D_{\ell_1}^{AB}} \Delta{D_{\ell_2}^{CD\,*}}}
  ({\cal M}_{\ell' \ell_2}^{CD\,-1})^{T}
\end{eqnarray}

The above equation as it stands can not be used in practice for $\ell$
above $\sim10$ as the calculation of
\begin{eqnarray}
\VEV{\Delta{D_{\ell_1}^{AB}} \Delta{D_{\ell_2}^{CD\,*}}} = 
  \frac{1}{(2\ell+1)(2\ell'+1)}
  \sum_{mm'} \sum_{\ell_1 m_1} \sum_{\ell_2 m_2}  \nonumber \\
  C_{\ell_1}^{AC} C_{\ell_2}^{BD} K_{\ell m\ell_1m_1}^A K_{\ell'm'\ell_1m_1}^C
  K_{\ell m\ell_2m_2}^B K_{\ell'm'\ell_2m_2}^D
\label{pseudocrosscorrelation}
\end{eqnarray}
described in Appendix~\ref{appendix_xi}, is numerically unstable
\citep{angularmomentum}. However, for high multipoles and sufficiently
large sky coverage (as it is the case for satellite missions as WMAP
and {\sc Planck}) it can be simplified \citep{efstathiou} by replacing
$ C_{\ell_1}^{AC}$ and $C_{\ell_2}^{BD}$ by $ C_{\ell^{\prime}}^{AC}$
and $C_{\ell^{\prime}}^{BD}$ respectively and then applying the
completeness relation for spherical harmonics \citep{angularmomentum}.
This is because the $K_{\ell m\ell'm'}$, which is diagonal for full
sky coverage, is quasi-diagonal for large sky coverage.
\\

Under the above hypothesis and following Appendix~\ref{appendix_xi},
the cross-correlation matrix reads
\begin{eqnarray}
  \Xi^{AB,CD}_{\ell \ell'} 
  & = &
  {\cal M}_{\ell \ell_1}^{AB\,-1}
  \left[
    \frac{
      {\cal M}^{(2)}_{\ell_1 \ell_2}\left(W^{AC,BD}\right) C_{\ell_1}^{AC} C_{\ell_2}^{BD}
    }{2\ell_2+1}
  \right.
  +
  \nonumber \\
  & & 
  \left.
    \frac{
      {\cal M}^{(2)}_{\ell_1 \ell_2}\left(W^{AD,BC}\right) C_{\ell_1}^{AD} C_{\ell_2}^{BC}
    }{2\ell_2+1}
  \right]
  ({\cal M}_{\ell' \ell_2}^{CD\,-1})^{T}
\label{xi}
\end{eqnarray}
where $${\cal M}_{\l1\l2}^{(2)}\left(W^{AC,BD}\right) = E_{\l1}^A
E_{\l1}^C M^{(2)}_{\l1 \l2}\left(W^{AC,BD}\right) E_{\l2}^B
E_{\l2}^D$$
and $M^{(2)}$ is the quadratic coupling kernel matrix

\begin{eqnarray}
  &&
  M^{(2)}_{\ell_1 \ell_2}\left(W^{AB,CD}\right) =
  \nonumber \\
  &&
  \frac{(2 \l2 + 1)}{4 \pi}
  \sum_{\l3} (2 \l3 + 1)
  W^{AB,CD}_{\ell_3} \wjjj{\l1}{\l2}{\l3}{0}{0}{0}^2
\end{eqnarray}
associated to the cross-power spectrum of the product of the masks $
W^{AB,CD} = \frac{1}{2\ell+1} \sum_{m}{w^{(2)}}^{AB}_{\ell m}
{w^{(2)}}^{CD*}_{\ell m}$. The ${w^{(2)}}^{AB}_{\ell m}$ represents the
spherical harmonic coefficients for the product of the masks
associated to detectors $A$ and $B$.
\\

Equation \ref{xi} can be further simplified by assuming uniform
weighting and the same sky coverage for all detectors, as well as a diagonal
dominated coupling kernel matrix,
\begin{equation}
  \Xi^{AB,CD}_{\ell \ell'} \simeq
  \frac{1}{\nu_{\ell'}}
  \left[
     C_\ell^{AC}C_{\ell'}^{BD} + C_\ell^{AD}C_{\ell'}^{BC}
  \right]
  \label{xi_approx}
\end{equation}

In this case, the effect of a non-homogeneous sky coverage is
represented by a simple function $\nu_\ell$ which can be associated to
the effective number of degrees of freedom in the $\chi^2_{\nu_\ell}$
distribution of the $\Cl$s over the sky (see \citealp{master})
\begin{equation}
  \nu_\ell = (2\ell+1) \Delta_\ell \frac{w_2^2}{w_4}
\end{equation}
where $w_i$ is the $i$-th moment of the mask
\begin{equation}
  w_i = \frac{1}{4\pi}\int_{4\pi} d\Omega W^i(\Omega) \, .
\end{equation}

In Eq.~\ref{xi} and \ref{xi_approx}, the $\Cl$ can be either
cross-power spectra or auto-power spectra depending on the combination
of $A$, $B$, $C$ and $D$ (the only condition is $A \neq B$ and $C \neq
D$).  In one hand, noise terms, as included in the auto-power spectra,
appear in the analytical form of the correlation matrix $\Xi$. On the
other hand, for a set of 4 independant detectors ($A \ne B \ne C \ne
D$), the correlation matrix is the variance of the signal.

\subsection[]{Covariance Matrix and error bars associated to the cross-power spectra}\label{covpercross}

It is important to notice that the cross-correlation matrix contains
all the needed information to perform error bars and covariance matrix
for each single cross-power spectrum. From Eq.~\ref{xi} (or
Eq.~\ref{xi_approx} for the approximated form), the covariance matrix
providing the correlation between adjacent multipoles is given by
\begin{eqnarray}
  Cov^{AB}(\ell,\ell')
  & = & 
  \Xi^{AB,AB}_{\ell \ell'} \nonumber \\
  & \simeq & \frac{1}{\nu_{\ell'}}
  \left[
    \Cl^{AB}C_{\ell'}^{AB} + \Cl^{AA} C_{\ell'}^{BB}
  \right]
  \label{covariance}
\end{eqnarray}

In the same way, we can also write the variance of each cross-power
spectra which corresponds to the error bars ${\Delta \Cl}^{AB}$
associated to each cross-power spectra
\begin{eqnarray}
  \left( {\Delta \Cl}^{AB} \right)^2
  & = & 
  \Xi^{AB,AB}_{\ell\ell} \nonumber \\
  & \simeq & \frac{1}{\nu_\ell}
  \left[
    \left(\Cl^{AB}\right)^2 + \Cl^{AA} \Cl^{BB}
  \right]
  \label{error_bar}
\end{eqnarray}

In these cases, instrumental noise appears in the auto-power spectra
$\Cl^{XX}$ (for which $N_\ell^{XX} \ne 0$). As it is computed directly
from the data, no estimation of the noise is needed.

\section[]{Combined angular power spectrum}\label{combination}
This section describes the estimation of the final CMB angular power
spectrum and the error bars associated to it using the corrected
cross-power spectra described in Sect.~\ref{cross} in addition to the
cross-correlation matrix computed in Sect.~\ref{correlation}. We
propose a simple but efficient way to combine them by maximizing a
quadratic likelihood function and to deduce the final error bars from
the cross-correlation matrix. For this paper we have considered a very
general form for the likelihood function limited to a simple CMB plus
noise model. However, the likelihood function can be adapted more
specifically to the data, for example to include point sources as it
is the case for the WMAP results presented in \citealp{wmap_cl}. More
generally, it could be extended to algorithms of spectral matching
decomposition ({\it e.g.}~\citealp{smica}). In the following, we present
how this method can be used to provide both an estimate of the $\Cl$s
from a full data set (Sect.~\ref{lincomb}) and of the angular power
spectrum of common sky fluctuations between maps coming from two or
more independent surveys (Sect.~\ref{common}).

\subsection[]{Gaussian approximated linear combination of the cross-power spectra}\label{lincomb}
Once all possible cross-power spectra have been computed from a data
set, we dispose of $N(N-1)/2$ different but not independent
measurements of the angular power spectrum, $C_\ell$s. To combine them
and obtain the best estimate of the power spectrum
$\widetilde{C_\ell}$, we maximize the Gaussian approximated likelihood
function
\begin{equation}
  -2 \ln {\cal L} = \sum_{ij} \left[ (C_\ell^i - \widetilde{\Cl}) |\Xi^{-1}|_{\ell\ell'}^{ij}
    (C_{\ell'}^j - \widetilde{C_{\ell'}}) \right]
\label{likelihood}
\end{equation} 
where $\left| \Xi_{\ell\ell'} \right|^{ij} = \Xi^{AB,CD}_{\ell\ell'}$
is the cross-correlation matrix of the cross-power spectra described
before ( $i$ and $j$ $\in \{AB, A \ne B \}$). The auto-power spectra
are not considered.

From this and neglecting the correlation between adjacent multipoles,
it is straightforward to show that the estimate of the angular power
spectrum is
\begin{equation}
  \widetilde{\Cl} = \frac{1}{2}
  \frac{ \sum_{ij} \left[ |\Xi^{-1}|_{\ell\ell}^{ij} C_\ell^j + C_\ell^i
      |\Xi^{-1}|_{\ell\ell}^{ij} \right]}
  { \sum_{ij} |\Xi^{-1}|_{\ell\ell}^{ij}}
\label{bestcell}
\end{equation}

The final covariance matrix can be obtained from Eq.~\ref{likelihood},
\begin{equation}
  Cov(\ell,\ell') = \frac{1}{\sum_{ij} |\Xi^{-1}|_{\ell\ell'}^{ij}}
\label{covbestcell}
\end{equation} 

and the final error bars are given by
\begin{equation}
  \left( \Delta \widetilde{\Cl} \right)^2 = \frac{1}{\sum_{ij}
    |\Xi^{-1}|_{\ell\ell}^{ij}} \, \, .
\label{errorbarbestcell}
\end{equation} 

Depending on the cross-correlation between cross-power spectra, the
instrumental noise variance is reduced by a factor comprised between
N, the number of independant detectors, and $N(N-1)/2$, the number of
cross-power spectra.

For the noise dominated case, the correlation between different
cross-power spectra can be neglected and the cross-correlation matrix
$\Xi$ becomes diagonal (see eq.~\ref{xi} and \ref{xi_approx}). The
values in the diagonal are the $N(N-1)/2$ products of 2 auto-power
spectra (including noise). The variance of final power spectrum is
then proportional to $\frac{1}{N(N-1)/2}$. In any case, when combining
cross-power spectra, the upper limit for the variance comes from the
combination of $N$ independent detectors (proportional to
$\frac{1}{N}$).

\subsection[]{Common angular power spectrum}\label{common}
The Xspect formalism allows us to compare two or more different sets
of sky maps coming from two or more independent experiments. In this
respect, the quantity of interest is the power spectrum of the common
fluctuations on the sky with the same physical origin. We will call
the latter common angular power spectrum $C_\ell^{common}$ for
simplicity. In fact, if we compare for example two sets of CMB maps at
low and high frequencies, the foreground contamination will be
different and we can expect to obtain a better estimation of the CMB
power spectrum. In the same way, for two different experiments,
systematic effects will be decorrelated and will not contribute to the
common angular power spectrum calculated from them. In addition,
template maps of foreground emission can be correlated to the CMB
experiments maps to monitor and to subtract foreground residuals.
\\

To simplify the notation we will only consider two sets of sky maps
$A$ and $B$, corresponding to $N_{A}$ and $N_{B}$ detectors
respectively. Following the previous paragraph considerations, by
cross-correlating each detector map from $A$ which each detector map
from $B$ and correcting the `pseudo' cross-power spectra as described
before, we can form $N_{A} \times N_{B}$ cross-power spectra which are
unbiased estimates of the common power spectrum but not fully
independent.

The cross-power spectra formed in this way can be noted
$C_{\ell}^{A_{i},B_{j}}$ and their cross-correlation matrix
$\Xi^{A_{i}B_{j},A_{u}B_{v}}_{\ell \ell'}$ reads
\begin{eqnarray}
  &&
  \Xi^{A_{i}B_{j},A_{u}B_{v}}_{\ell \ell'}=
  \nonumber \\
  &&
  {\cal M}_{\ell \ell_1}^{A_{i}B_{j}\,-1}
  \VEV{\Delta{D_{\ell_1}^{A_{i}B_{j}}} \Delta{D_{\ell_2}^{A_{u}B_{v}\,*}}}
  ({\cal M}_{\ell' \ell_2}^{A_{u}B_{v}\,-1})^{T}
\end{eqnarray}
which can be computed using previous approximations. \\

Finally, equations \ref{bestcell} and \ref{errorbarbestcell} are used
to obtain the estimate of the common angular power spectrum
$\widetilde{C_\ell}^{common}$ and the error bars associated to it.

This method allows us to deal naturally with very different
experimental configurations such as for example completely different
scanning strategies, different timeline filtering and different beam
patterns for each of the detectors involved. Furthermore, it can be
easily generalized to work with non CMB data such as maps of the
Sunyaev-Zel'dovich effect in clusters of Galaxies or maps of mass
fluctuations from weak lensing observations.

\section[]{Xspect applied to Archeops balloon-borne experiment}\label{archeops}

Xspect was mainly developed to measure the CMB temperature angular
power spectrum from the {\sc Archeops} data as well as to compare
these data to other observations of the sky as for example those from
the WMAP satellite. {\sc Archeops} \citep{archeops_cl} is a
balloon-borne experiment conceived as a precursor of the {\sc Planck}
High Frequency Instrument\footnote{http://www.planck-hfi.org}. It
consists of a 1.5~m off-axis Gregorian telescope and of an array of
21~photometers cooled down to $\sim 100$~mK and which operates in
4~frequency bands: 143 and 217 GHz (CMB channels) and 353 and 545 GHz
(Galactic channels). For the latest flight campaign, the entire {\sc
  Archeops} data cover about $\sim 30$\% of the sky, including the
Galactic plane.

\subsection[]{Simulations of the Archeops data set}
As a first step in the application of Xspect to the {\sc Archeops}
data set, we have produced 500 simulations of the sky observed by {\sc
  Archeops} including both the CMB signal and the detector noise for
the six most sensitive {\sc Archeops} detectors at 143 and 217 GHz.

These simulations are computed from realizations of the CMB sky at
$nside=512$ (corresponding to a pixel size of $\sim 7$~arcmin) for the
{\sc Archeops} best-fit CMB model presented in \citealp{archeops_cosmo}.
The sky maps are convolved by the main beam pattern of each of the
\mbox{$N_{detec}=6$}~{\sc Archeops} detectors using the beam transfer
function and then deprojected following the {\sc Archeops} scanning
strategy to produce mock {\sc Archeops} timelines. The beam transfer
functions were computed individually for each detector from Jupiter's
crossings in the data using the {\it Asymfast} method presented in
\citealp{asymfast}.

\begin{figure}
  \includegraphics[scale=0.5]{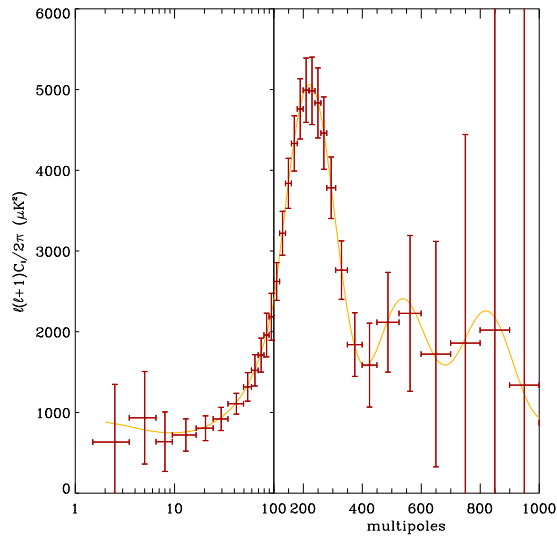} 
  \caption{
    Mean angular power spectrum obtained from $500$ simulations of
    {\sc Archeops} observations of the CMB sky (see text for details).
    We use the {\sc Archeops} best-fit model \citep{archeops_cl} to
    produce CMB maps which are smeared out using the beam pattern of
    each of the six most sensitive {\sc Archeops} detectors and
    deprojected into mock {\sc Archeops} timelines. Noise is then
    added up for each detector and the resulting timelines projected
    into sky maps are analyzed using Xspect. The error bars shown here
    are computed analytically as described in previous sections. The
    Xspect estimate of the angular power spectrum (in red) is an
    unbiased estimate of the input CMB model (in yellow).}
  \label{cl_simuarcheops}
\end{figure}

The detector noise is computed from the time power spectrum of each of
the detectors and added to the mock signal timelines. The method used
for the estimation of the time noise power spectra is described in
details in \citealp{noise_estimator}. To suppress $1/f$-noise, remaining
atmospheric and galactic contamination as well as non-stationary high
frequency noise, the {\sc Archeops} time streams are Fourier filtered
 using a bandpass filter between 0.1 and
38~Hz. We apply the same filtering to the simulated timelines before
projection. The corresponding multipole filtering function $F_{\ell}$
was independently calculated by specific simulations. The filtered
timelines are then reprojected using the {\sc Archeops} scanning
strategy to produce $N_{detec}$ simulated co-added maps for each
simulation run.

A Galactic mask deduced from a SFD IRAS map \citep{schlegel},
extrapolated to 353~GHz, using a cut in amplitude (greater than
0.5~MJy.str) is applied to the simulated maps. This reduces the total
{\sc Archeops} coverage to $\sim 20\%$ of the sky. We have used two
different weighting schemes on the maps. The first one is an uniform
weighting and the second one, a $1/\sigma_{pix}^2$ weighting per pixel
and per detector, where $\sigma_{pix}^2$ is the noise variance for the
given pixel.

Then we apply the Xspect method to each simulations-run. The
$N_{detec}(N_{detec}-1)/2$ cross-power spectra are computed with their
associated cross-correlation matrices and combined as
described in Sect.~\ref{lincomb} using the Gaussian approximation of
the likelihood function.

In addition to the above simulations, we produced, in the same way,
500 non noisy simulations from which we can extract for example the
sample variance for the {\sc Archeops} coverage.

Figure~\ref{cl_simuarcheops} shows the mean angular power spectrum
computed from the angular power spectrum estimates obtained for each
of the 500-simulations runs. The error bars are analytically computed.
We observe that the Gaussian approximated combination of cross-power
spectra is an unbiased estimate of the input CMB model for the angular
power spectrum. This mean angular power spectrum is a combination of
two weighting schemes. Up to $\ell=90$, we use uniform weighting
whereas, for larger $\ell$, a mask inversely proportional to the noise
in each of the pixels has been applied. This allows us to choose the
smaller error bars in each region.

\begin{figure}
  \includegraphics[scale=0.5]{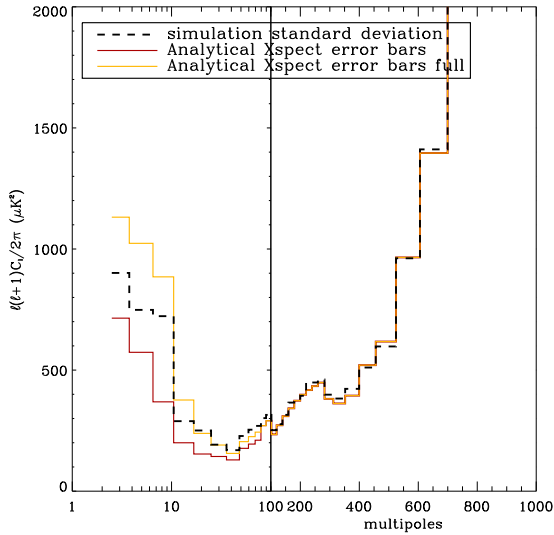}
  \caption{\label{error_simuarcheops} 
    Error bars for the {\sc Archeops} data computed analytically, from
    Eq.~\ref{xi_approx} (red solid line) and from Eq.~\ref{xi} (yellow
    solid line). They are compared to the standard deviation of
    $N_{simu}=500$ simulations (black dashed line). Differences
    essentially originate from the extremely inhomogeneous {\sc
      Archeops} sky coverage (see text for details).}
\end{figure}

\begin{figure}
  \includegraphics[scale=0.3,angle=90]{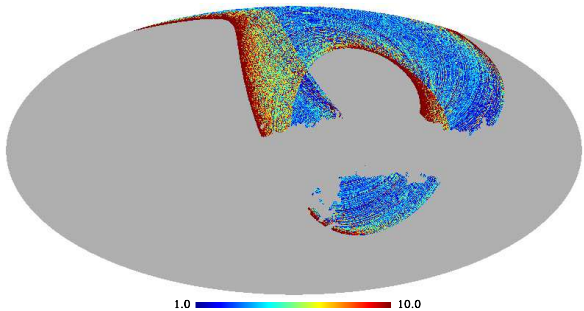}
  \caption{\label{archeops_cover}
    Sky coverage and weighting scheme applied to the sky map
    ($nside~=512$) obtained from the {\sc Archeops} most sensitive
    detector at 143~GHz. We observe a ring-like structure with a large
    uncovered area in the center and highly redundant areas on the
    edges of the ring.}
\end{figure}

Figure~\ref{error_simuarcheops} shows analytic estimates of the error
bars in the final angular power spectrum compared to the dispersion
for each multipole bin over the 500 Monte-Carlo simulations. The {\sc
  Archeops} sky coverage is very inhomogeneous due to the particular
choice of the scanning strategy which tries to maximize the area of
the sky observed ($\sim 30$\%) in a very reduced amount of
observation time (less than 12~hours). {\sc Archeops} performs large
circles on the sky that leads to a quasi ring-like coverage with a
large uncovered region at the center of the ring as shown on
Fig.~\ref{archeops_cover}. This implies that the approximation used to
obtain Eq.~\ref{xi_approx} is not valid, especially at low multipoles,
and therefore the error bars obtained analytically (red solid line)
are underestimated up to $\sim 20\%$ at very low $\ell$ with a mean of
$\sim 10\%$. By computing $\Xi^{AB,CD}_{\ell\ell'}$ as given by
Eq.~\ref{xi}, the analytic error bars (yellow solid line) increase as
expected and fit much better the dispersion in the simulations (black
dashed line). The agreement is then within $10\%$ over the full range
of multipoles with a mean of $1.5\%$. Equivalent results are obtained
using non noisy simulations.

\subsection[]{Application to the {\sc Archeops} data}

Xspect has been applied to the data from the last {\sc Archeops}
flight campaign using the six most sensitive bolometers as in the
simulations presented above. The results of this analysis is presented
in a collaboration paper \citep{archeops_cl2}. Cross-correlation with
WMAP maps is also under study in order to assess the electromagnetic
spectrum of the CMB anisotropies from 40 to 217~GHz.

\section[]{Summary and Conclusions}
In this paper, we have presented a method, Xspect, for the obtention
of the CMB angular power spectrum with analytical error bars based on
four main steps:
\begin{enumerate}
\item Given $N$ independent input sky maps from different detectors
  either from a single experiment or from multiple ones, we estimate
  $N(N-1)/2$ `pseudo' cross-power spectra as well as $N$ `pseudo'
  auto-power spectra using the HEALPix package \citep{healpix}.
\item Cross-power spectra are obtained by correcting the `pseudo'
  power spectra for weighting scheme, beam smoothing and filtering as
  discussed in Section~\ref{cross}. In the same way, we compute
  auto-power spectra with noise. Xspect can deal with different and
  complex weighting schemes for each of the detectors involved. The
  coupling matrix, beam and filtering transfer functions are
  precomputed for each pair of detectors.
\item We compute the full cross-correlation matrix, $\Xi^{AB,CD}_{\ell
    \ell'}$, between cross-power spectra $AB$ ($A \ne B$) and $CD$ ($C
  \ne D$) (Section~\ref{correlation_matrix}) from which we can extract
  the covariance matrix and the error bars for each cross-power
  spectra (Sect~\ref{covpercross}).
\item Finally, the corrected cross-power spectra are combined into a
single angular power spectrum using their cross-correlation matrices
which are assumed to be diagonal in multipole space. The Gaussian
approximation of the likelihood function is fully justified in the
large multipole range. In addition, analytical estimates for the error
bars and for the covariance matrix are computed
(Section~\ref{combination}). The covariance matrix can be used to
check the degree of correlation between multipoles.
\end{enumerate}

As this method estimates the angular power spectrum using `pseudo'
cross-power spectra, we can obtain a non noise-biased power spectrum
avoiding the estimation of the noise power spectra which requires
heavy Monte-Carlo simulations. Equally, this permits the estimation of
analytical error bars which are compatible to those computed from
simulations. Furthermore, Xspect allows us to obtain the angular power
spectrum of common sky fluctuations between two or more experiments.
Associated to other surveys used as templates, it can provide
estimations of systematic residuals or astrophysical contaminations.
Nevertheless, Xspect computes only the common structures between maps,
assuming a single physical component. This means that, for CMB
purposes, foregrounds and systematics must be subtracted beforehand.

Xspect has been successfully applied to simulations of the {\sc
  Archeops} experiment which presents an inhomogeneous sky coverage
and detectors with unequal sensitivities. Even for a balloon-borne
experiment that can be compared to satellite missions neither in noise
level nor in sky coverage, the analytical estimates of the errors
computed with Xspect are accurate within few percents. This property
is of great interest when producing Monte Carlo simulations in the
process of testing and improving the estimation of the $\Cl$s or when
checking the robustness of the data analysis with respect to the
various choices of mask, filtering or binning. The application of
Xspect to the {\sc Archeops} data set will be published by the {\sc
  Archeops} collaboration shortly \citep{archeops_cl2}.

For the WMAP satellite, which covers the full sky in a very
homogeneous way, it was also interesting to check the analytical
estimates of the error bars. Xspect analytical errors are equivalent
to those provided by the WMAP team within 15\% with a mean of
1.8\%. This agreement is satisfactory as our analysis was more
basic than the one used to derive the first year WMAP results
presented in \citealp{wmap_cl}. In particular all maps at all multipoles
were used without any point-source specific treatment and only two
weighting schemes were applied.

The extension of Xspect to CMB polarization maps is under development.
As for the temperature power spectrum, the $C_{\ell}^{EE}$ and
$C_{\ell}^{BB}$ power spectra obtained directly from a single set of
I, Q and U maps via a 'pseudo' power spectrum estimator are noise
biased. A 'pseudo' cross-power spectrum estimator adapted to
polarization can solve this problem by using independent sets of I, Q
and U maps.

\section*{Acknowledgments}
  Authors are grateful to J-Ch.~Hamilton for discussions in the
  initial stage of this work in our team. We also want to thank
  F.X~D\'esert for useful discussions on the method and for
  corrections of this paper. The Healpix package \citep{healpix} was
  used extensively .

\onecolumn
 


\onecolumn
\newpage
\appendix
\section[]{correlation matrix}\label{appendix_xi}
In this appendix we describe in details the calculation of the
cross-correlation matrix $\Xi^{AB,CD}_{\ell \ell'}$ under the hypothesis of large sky coverage
which leads to

\begin{eqnarray}
  \Xi^{AB,CD}_{\ell \ell'}
  & \equiv  &
  \VEV{ \Delta C_\ell^{AB} \Delta C_{\ell'}^{CD^*} }
  \\
  & = &
  {\cal M}_{\ell \ell_1}^{AB\,-1}
  \VEV{\Delta{D_{\ell_1}^{AB}} \Delta{D_{\ell_2}^{CD\,^*}}}
  ({\cal M}_{\ell' \ell_2}^{CD\,-1})^{T} \mbox{\,\,where\,\,} \Delta X = \hat{X} - \VEV{X}
\label{appcrosscorr}
\end{eqnarray}

Let us remind that for Gaussian distributed variables $x_i$ with
variance $\left< x_i x_j \right> = \sigma_{ij}^2$ the quadri-variance
is given by
\begin{eqnarray}
  \label{4var}
  \left< x_i x_j x_k x_l \right>
  & = &
  \left< x_i x_j \right> \left< x_k x_l \right> +
  \left< x_i x_k \right> \left< x_j x_l \right> +
  \left< x_i x_l \right> \left< x_j x_k \right> \nonumber \\
  & = &
  \sigma_{ij}^2 \sigma_{kl}^2 + \sigma_{ik}^2\sigma_{jl}^2 +
  \sigma_{il}^2\sigma_{jk}^2
\end{eqnarray}

Using Eq~\ref{4var}, the main term of the cross-correlation matrix in 
Eq.~\ref{appcrosscorr} reads
\begin{eqnarray}
  \VEV{ \widehat{D_{\ell_1}^{AB}} \widehat{D_{\ell_2}^{CD}}^*}
  & = &
  \frac{1}{(2\ell_1+1)(2\ell_2+1)}
  \sum_{m_1m_2}
  \VEV{ d_{\ell_1m_1}^A d_{\ell_1m_1}^{B*} d_{\ell_2 m_2}^{C*} d_{\ell_2 m_2}^D }
  \nonumber \\
  & = &
  \sum_{m_1m_2}
  \frac{
    \VEV{d_{\ell_1 m_1}^A d_{\ell_1 m_1}^{B*}}
    \VEV{ d_{\ell_2 m_2}^{C*} d_{\ell_2 m_2}^D} +
    \VEV{d_{\ell_1 m_1}^A d_{\ell_2 m_2}^{C*}}
    \VEV{ d_{\ell_1 m_1}^{B*} d_{\ell_2 m_2}^D} +
    \VEV{d_{\ell_1 m_1}^A d_{\ell_2 m_2}^D}
    \VEV{d_{\ell_1 m_1}^{B*} d_{\ell_2 m_2}^{C*}}
  }{(2\ell_1+1)(2\ell_2+1)}
  \nonumber \\
  & = &
  D_{\l1}^{AB} D_{\l2}^{CD} +
  \frac{1}{(2\l1+1)(2\l2+1)}
  \sum_{\m1\m2}
  \left[
    \VEV{d_{\l1\m1}^A d_{\l2\m2}^{C*}} \VEV{d_{\l1\m1}^{B*} d_{\l2\m2}^D} +
    \VEV{d_{\l1\m1}^A d_{\l2\m2}^D} \VEV{d_{\l1\m1}^{B*} d_{\l2\m2}^{C*}}
  \right]
  \nonumber
\end{eqnarray}

where $\dlm$ are the coefficients of the spherical harmonic decomposition of
the masked sky map such that
\begin{equation}
  \dlm =
  \sum_{\l1\m1} a_{\l1\m1}
K_{{\ell m}{\l1\m1}}
\end{equation}
with
\begin{eqnarray}
  K_{{\ell_1 m_1}{\ell_2 m_2}}
  & = &
  \sum_i w_i \Omega_i Y_{\ell_1m_1}({\bf \theta}_i) Y_{\ell_2m_2}^{*}({\bf \theta}_i)
  \\
  & = &
  \sum_{\ell_3 m_3} (-1)^{m_2} w_{\ell_3 m_3}
  \left ( \frac{(2 \ell_1 + 1) (2 \ell_2 + 1) (2 \ell_3 + 1)}{4 \pi} \right )^{1/2}
  \wjjj{\ell_1}{\ell_2}{\ell_3}{0}{0}{0}
  \wjjj{\ell_1}{\ell_2}{\ell_3}{m_1}{-m_2}{m_3}
\end{eqnarray}
\noindent $\Omega_i$ is the area of pixel $i$. $w_i$ represents the
mask applied to the sky map and $w_{\ell m}$ the coefficients of the
spherical harmonic decomposition of the mask
\begin{equation}
 w_{\ell m} = \sum_i w_i \Omega_i Y_{\ell m}({\bf \theta}_i).
\end{equation}

Thus, the cross-correlation matrix can be simply written as follows
\begin{eqnarray}
  \VEV{\Delta{D_{\l1}^{AB}} \Delta{D_{\l2}}^{CD\,^*}}
  & = &
  \frac{1}{(2\l1+1)(2\l2+1)}
  \sum_{\m1\m2}
  \left[
    \VEV{d_{\l1\m1}^A d_{\l2\m2}^{C*}} \VEV{ d_{\l1\m1}^{B*} d_{\l2\m2}^D} +
    \VEV{d_{\l1\m1}^A d_{\l2\m2}^{D}} \VEV{ d_{\l1\m1}^{B*} d_{\l2\m2}^{C*}}
  \right] \label{delta_var}
\end{eqnarray}

Let us develop each of the terms in Eq. \ref{delta_var}, using $E_\ell =p_\ell B_\ell \sqrt{F_\ell}$
\begin{eqnarray}
  \VEV{\dlm^X d_{\ell'm'}^{Y*}}
  & = &
  \sum_{\ell_1 m_1} \sum_{\ell_2 m_2}
  \VEV{a_{\ell_1m_1}^X a_{\ell_2m_2}^{Y*}} E_{\l1}^X E_{\l2}^Y K_{\ell m\ell_1 m_1}^X K_{\ell'm'\ell_2m_2}^{Y*}
  \nonumber \\
  & = &
  \sum_{\ell_1 m_1} C_{\ell_1}^{XY} E_{\l1}^X E_{\l1}^Y K_{\ell m\ell_1m_1}^X K_{\ell_1m_1\ell'm'}^Y
  \nonumber \\
  & = &
  \sum_{\ell_1 m_1} C_{\ell_1}^{XY} E_{\l1}^X E_{\l1}^Y \sum_{ij} w^X_i w^Y_j
  \Omega_i \Omega_j
  Y_{\ell m}({\bf \theta}_i) Y_{\ell_1 m_1}^*({\bf \theta}_i)
  Y_{\ell' m'}^*({\bf \theta}_j) Y_{\ell_1 m_1}({\bf \theta}_j) 
  \nonumber 
\end{eqnarray}

\noindent Assuming a large sky coverage \citep{efstathiou} we can replace 
$C_{\ell_1}^{XY}$ by $C_{\ell}^{XY}$ and then, applying the
completeness relation for spherical harmonics $\sum_{\ell m} Y_{\ell
  m}(\theta_i) Y_{\ell m}^*(\theta_j) = \frac{1}{\Omega_i}\delta(\theta_i-\theta_j)$, we
obtain
\begin{eqnarray}
  \VEV{\dlm^X d_{\ell'm'}^{Y*}}
  & = &
  C_{\ell}^{XY} E_\ell^X E_\ell^Y \sum_{i} \left(w^X_i w^Y_i\right)
  \Omega_i
  Y_{\ell m}({\bf \theta}_i) Y_{\ell' m'}^*({\bf \theta}_i)
  \nonumber \\
  & = &
  C_{\ell}^{XY} E_\ell^X E_\ell^Y K^{(2)\,XY}_{\ell m\ell'm'}
\end{eqnarray}

Thus, under the above hypothesis Eq.~\ref{delta_var} reads 
\begin{eqnarray}
  \VEV{\Delta{D_{\l1}^{AB}} \Delta{D_{\l2}^{CD\,^*}}}
  & = &
  \frac{E_{\l1}^A E_{\l1}^C E_{\l2}^B E_{\l2}^D}{(2\l1+1)(2\l2+1)}
  \sum_{\m1\m2} \left[
    K^{(2)\,AC}_{\l1\m1\l2\m2} K^{(2)\,BD}_{\l1\m1\l2\m2} C_{\l1}^{AC} C_{\l2}^{BD} + 
    K^{(2)\,AD}_{\l1\m1\l2\m2} K^{(2)\,BC}_{\l1\m1\l2\m2} C_{\l1}^{AD} C_{\l2}^{BC}
  \right]
  \label{one_term}
\end{eqnarray}

\noindent The above expression is equivalent to that for the coupling
matrix $\Mll$ presented in \citealp{master} and can be simplified as
follows
\begin{eqnarray}
  \sum_{mm'}
  K^{(2)\,AC}_{\ell m\ell'm'} K^{(2)\,BD}_{\ell m\ell'm'}
  & = &
  \sum_{mm'} \sum_{\ell_1m_1} \sum_{\ell_2m_2}
  w^{AC}_{\ell_1m_1} w^{BD}_{\ell_2m_2}
  \frac{(2\ell+1)(2\ell'+1)}{4\pi} 
  \left( (2\ell_1+1)(2\ell_2+1) \right)^{1/2}
  (-1)^{m_1+m_2}
  \nonumber \\
  & &
  \wjjj{\ell}{\ell'}{\ell_1}{0}{0}{0}\wjjj{\ell}{\ell'}{\ell_2}{0}{0}{0}
  \wjjj{\ell}{\ell'}{\ell_1}{m}{-m'}{-m_1}\wjjj{\ell}{\ell'}{\ell_2}{m}{-m'}{-m_2}
  \nonumber \\
  & = &
  (2\ell+1) M^{(2)}_{\ell\ell'}\left(W^{AC,BD}\right)
  \label{quadratic_mll}
\end{eqnarray}

\noindent where $M^{(2)}$ is the quadratic coupling kernel matrix 
\begin{equation}
  M^{(2)}_{\ell_1 \ell_2}\left(W^{AB,CD}\right)
  =
  (2 \ell_2 + 1)
  \sum_{\ell_3 } \frac{(2 \ell_3 + 1)}{4 \pi}
  W^{AB,CD}_{\ell_3}
  {\left ( \begin{array}{ccc}
        \ell_1 & \ell_2 & \ell_3  \\
        0  & 0 & 0
      \end{array}
    \right )^2}
\end{equation}
\noindent associated to
the cross-power spectrum of the products of the masks
$w^{XY}_{i}=w^{X}_{i}*w^{Y}_{i}$ for each of the cross-power spectra
$AB$ et $CD$,
$$W^{AB,CD} = \frac{1}{2\ell+1} \sum_{m}w^{AB}_{\ell m}w^{CD*}_{\ell m}$$

\noindent Replacing \ref{quadratic_mll} in \ref{one_term} we obtain the following
expression
\begin{eqnarray}
  \VEV{ \Delta D_{\ell_1}^{AB} \Delta D_{\ell_2}^{CD\,*}}
  & = &
  \frac{E_{\l1}^A E_{\l1}^C E_{\l2}^B E_{\l2}^D}{2\ell_2+1}
  \left[
    M^{(2)}_{\ell_1 \ell_2}\left(W^{AC,BD}\right) C_{\ell_1}^{AC} C_{\ell_2}^{BD}
    +
    M^{(2)}_{\ell_1 \ell_2}\left(W^{AD,BC}\right) C_{\ell_1}^{AD} C_{\ell_2}^{BC}
  \right]
\end{eqnarray}

\noindent and thus, using the same abreviation ${\cal M}_{\l1\l2}^{(2)}\left(W^{AC,BD}\right) = E_{\l1}^A E_{\l1}^C
M^{(2)}_{\ell_1 \ell_2}\left(W^{AC,BD}\right) E_{\l2}^B E_{\l2}^D$ the cross-correlation matrix reads,
\begin{eqnarray}
  \Xi^{AB,CD}_{\ell \ell'}
  & = &
  {\cal M}_{\ell \ell_1}^{AB\,-1}
  \left[
    \frac{
      {\cal M}^{(2)}_{\ell_1 \ell_2}\left(W^{AC,BD}\right) C_{\ell_1}^{AC} C_{\ell_2}^{BD}
      +
      {\cal M}^{(2)}_{\ell_1 \ell_2}\left(W^{AD,BC}\right) C_{\ell_1}^{AD} C_{\ell_2}^{BC}
    }{2\ell_2+1}
  \right]
  ({\cal M}_{\ell' \ell_2}^{CD\,-1})^{T}
\end{eqnarray}

\end{document}